\documentclass[twocolumn,showpacs,preprintnumbers,amsmath,amssymb]{revtex4}

\usepackage{graphicx}
\usepackage{color}
\usepackage{dcolumn}
\usepackage{bm}

\begin{document}

\preprint{APS/123-QED}

\title{Dynamics and correlation length scales of a glass-forming liquid in quiescent and sheared conditions}

\author{Wen-Sheng Xu, Zhao-Yan Sun\footnote{Correspondence author. E-mail: zysun@ciac.jl.cn}, and Li-Jia An\footnote{Correspondence author. E-mail: ljan@ciac.jl.cn}}
\affiliation{State Key Laboratory of Polymer Physics and Chemistry,
Changchun Institute of Applied Chemistry, Chinese Academy of
Sciences, Changchun 130022, People's Republic of China}



\date{\today}

\begin{abstract}
We numerically study dynamics and correlation length scales of a colloidal liquid in both quiescent and sheared conditions to further understand the origin of slow dynamics and dynamic heterogeneity in glass-forming systems. The simulation is performed in a weakly
frustrated two-dimensional liquid, where locally preferred order is
allowed to develop with increasing density. The four-point density
correlations and bond-orientation correlations, which have been
frequently used to capture dynamic and static length scales
$\xi$ in a quiescent condition, can be readily extended to a system
under steady shear in this case. In the absence of shear, we
confirmed the previous findings that the dynamic slowing down
accompanies the development of dynamic heterogeneity. The dynamic
and static length scales increase with $\alpha$-relaxation time
$\tau_{\alpha}$ as power-law $\xi\sim\tau_{\alpha}^{\mu}$ with
$\mu>0$. In the presence of shear, both viscosity and
$\tau_{\alpha}$ have power-law dependence on shear rate in the
marked shear thinning regime. However, dependence of correlation
lengths cannot be described by power laws in the same regime. Furthermore,
the relation $\xi\sim\tau_{\alpha}^{\mu}$ between length scales and
dynamics holds for not too strong shear where thermal
fluctuations and external forces are both important in determining
the properties of dense liquids. Thus, our results demonstrate a link
between slow dynamics and structure in glass-forming liquids even
under nonequilibrium conditions.
\end{abstract}

\pacs{61.20.Ja, 61.20.Lc, 64.70.P-}

\maketitle

\section{Introduction}
Glass can be formed in a variety of molecular and colloidal systems, but the nature of the glass
transition remains elusive despite intensive
research~\cite{PWAnderson, Angell, Debenedetti, Berthier1}. It is
commonly observed that time scales, such as structural relaxation
time, dramatically increase on the approach to the glass transition
point with little change in the static structure of the liquid. This
is the reason why the research on the glass transition for a long
time has been focused on time scales much more than length scales.
In the past decade, however, both experiments and simulations have
revealed that the relaxation process not only drastically slows down
but also becomes progressively more heterogeneous as a liquid
approaches its glass transition point~\cite{Kegel, Weeks, Berthier2,
Kob, Glotzer1, Yamamoto1, Yamamoto2}. The collective nature of the
dynamics of the supercooled liquids can indeed date back to the
seminal idea of Adam and Gibbs~\cite{AdamGibbs} and has been well
reviewed in recent publications~\cite{Ediger, HCAndersen, Berthier1,
Berthier3, Berthier4}. The phenomenon of growing dynamic
heterogeneity approaching the glass transition point also suggests
the existence of a growing dynamic length scale in glass-forming
systems~\cite{Yamamoto2, Glotzer1, Glotzer2}. As a matter of fact,
the issue of characteristic length scales has become one of the main
themes of the study on the glass transition~\cite{Berthier1}.

To further understand the origin of the dynamic heterogeneity and
explore whether there exist structural signatures connected to the
slow dynamics and dynamic heterogeneity in viscous liquids, several
ideas have been put forward. One of the ideas is based on the effect
of frustration on crystallization in vitrification. To
continuously control frustration, several ways are frequently used
in the study~\cite{Tarjus, Tanaka5, Sausset2}. One way is to focus on the
polydisperse colloidal liquid where frustration can be continuously
changed by varying polydispersity~\cite{Tanaka5} (as inspired by the study of crystal nucleation on
polydisperse colloids~\cite{Frenkel}) and another way is to study
liquids in negatively curved space where frustration can be
controlled by varying space curvature~\cite{Sausset2} (as
motivated from the frustration-based approach of the glass
transition~\cite{Tarjus}). In both cases, if the frustration is not
very strong, some local order reminiscent of the crystal structure
can develop with increasing density or decreasing temperature. Then,
one can readily detect some static correlations (usually by bond-orientation
correlation functions) approaching the glass transition and
investigate whether these structural signatures are connected to the
dynamical properties. Although different interpretations were given
for the change of correlation length scales approaching the glass
transition point~\cite{Tanaka5, Sausset2}, the structural signature of slow dynamics and dynamic
heterogeneity does exist at least in these model
glass-forming liquids. In particular, a direct relation between
dynamic heterogeneity and the so-called medium-range crystalline
order (MRCO, i.e., long-lived clusters of particles with high
structural order) has been revealed in colloidal
dispersions~\cite{Tanaka5}. Another more
general method to detect the nontrivial static spatial correlations
has been proposed recently, which is motivated from the random first
order transition (RFOT) theory~\cite{Kirkpatrick1, Kirkpatrick2}
(see recent reviews~\cite{Biroli1, Berthier1} for details on this
theory). Within this theory, the so-called point-to-set
correlations~\cite{Biroli2, Berthier5, Kob1, Biroli3, Biroli4} are
believed to capture nontrivial structures of viscous liquids. The
idea is to first freeze the position of a set of particles in an
equilibrium configuration and then let the system evolve in the
presence of the constraint and measure how the position of the
remaining particles is affected. Thus, the point-to-set length is a
measure of the spatial extent over which the effect of equilibrium
amorphous boundary conditions propagates. Numerical simulations have
confirmed qualitatively the growth of the point-to-set
correlations~\cite{Biroli2} and its connection to the dynamic
correlations has also been explored recently~\cite{Berthier5, Kob1}.
Besides, the structural features of slow dynamics in viscous liquids
have also been investigated in Refs.~\cite{WidmerCooper1,
WidmerCooper2, WidmerCooper3, WidmerCooper4, Mittal2, Pedersen, Coslovich}.

In recent years, the study of supercooled liquids and glasses under
nonequilibrium conditions (e.g., under shear) has also attracted
considerable attention, with emphasis on their rheological properties and
the establishment of the concept of effective
temperature~\cite{Liu, Berthier6, Varnik1, Besseling1, Fuchs1}. The rheology of soft
glassy materials not only provide new insight into the nature of the
glass transition but also has its own value. One common observation
is the marked shear-thinning behavior in glass-forming liquids. In
particular, the shear viscosity $\eta$ is found to decrease with
increasing the shear rate $\dot{\gamma}$ as $\eta \sim
\dot{\gamma}^{-\nu}$ with $\nu\leq1$ . Another intriguing finding is
that soft glassy materials commonly display the shear banding
phenomenon~\cite{Varnik1}, i.e., a system under strong
shear spontaneously ``phase separates'' between a flowing region
supporting the shear and an immobile region with no flow. Recently,
the question of how dynamic heterogeneity is influenced by the flow has also been addressed~\cite{Bocquet2, Furukawa, Heussinger1, Heussinger2, Tsamados, Nordstrom, Mizuno}, e.g., numerical simulations~\cite{Furukawa} revealed the
existence of large scale heterogeneities under shear flow and its connection to the non-Newtonian behavior of a
supercooled liquid has also been explored, and power-law dependence of the intensity of the dynamic heterogeneity on shear rate has been observed~\cite{Heussinger1, Heussinger2, Tsamados, Nordstrom, Mizuno}. However, it is still not clear how dynamic
heterogeneity, MRCO and the associated length scales respond to shear.

In this work, we focus on slow dynamics and dynamic heterogeneity of
a colloidal glass-forming liquid and their structural features in
both quiescent and sheared conditions. We use a weakly frustrated
two-dimensional (2D) liquid as our model system, where
crystallization can be avoided and locally preferred order is
allowed to develop with increasing density. The four-point density
correlations and bond-orientation correlations, which have been
frequently used to capture the dynamic and static length scales
$\xi$ in a quiescent condition, can be readily extended to a system
under steady shear in this case. We confirmed the previous findings
in the absence of shear that slow dynamics accompanies the
development of dynamic heterogeneity. Both dynamic and static length
scales increase with $\alpha$-relaxation time $\tau_{\alpha}$ as power-law $\xi\sim\tau_{\alpha}^{\mu}$ with $\mu>0$. In the presence
of shear, both $\eta$ and $\tau_{\alpha}$ have power-law dependence
on $\dot{\gamma}$ in the marked shear thinning regime. However, dependence of the dynamic
and static correlation lengths on $\dot{\gamma}$ cannot be
described by power laws in the same regime. We further find that
the relation $\xi\sim\tau_{\alpha}^{\mu}$ between length scales and
dynamics holds for not too strong shear where thermal
fluctuations and applied forces are both important in determining
the properties of dense liquids. The paper is organized as follows: In
Sec. II, we show the model and methods used in this study, and then we
present results and discuss structural signatures of slows dynamics
and dynamic heterogeneity in the absence of shear in Sec. III.
Results under steady shear are given in Sec. IV. Finally, Sec. V
summarizes our results.

\section{Model and Methods}

The model we focus here is a 2D liquid, in which the particles
interact via the Weeks-Chandler-Andersen (WCA) potential~\cite{WCA}:
\begin{equation}
U_{jk}(r)=\left\{\!\!\!
\begin{array}{ll}
4\epsilon [(\sigma_{jk}/r)^{12}-(\sigma_{jk}/r)^{6}+1/4], & \text{for $r<2^{1/6}\sigma_{jk}$}\\
0, &
\text{otherwise,}
\end{array}\right.
\end{equation}
where $\epsilon$ is the depth of
the potential well, $r$ is the distance between two particles, and
$\sigma_{jk}=(\sigma_{j}+\sigma_{k})/2$ with $\sigma_{j}$ the
diameter of particle $j$. As the WCA potential is short-ranged
repulsive, a WCA liquid at low temperatures behaves as a colloidal liquid.
Although recent simulations~\cite{Attract1, Attract2} have shown that the longer-ranged attractive interactions can affect the dynamics of viscous liquids in quantitatively and qualitatively ways, we expect that the conclusions in our work will not be changed since the purely repulsive model can also capture the essential features of a typical glass-forming liquid.
To introduce weak frustration on crystallization in vitrification, the particle
diameters were chosen equidistantly from the range $0.8-1.2$ with an
interval of $0.01$, then the polydispersity for the system is
$\Delta=\sqrt{(<\sigma^{2}>-<\sigma>^{2})}/<\sigma>=11.98\%$, where
$<x>$ is the average of variable $x_{j}$ among all the
particles. Such polydispersity can avoid crystallization in the
system and allow static structural order to develop with increasing
density. In principle, Brownian dynamics simulation is a more
faithful method to study colloids. However, previous work has
suggested that at long times, when particles have collided many
times, Brownian and Newtonian systems show qualitatively the same
phenomenology~\cite{Lowen}. We employed molecular dynamics
simulation in the $NVT$ ensemble, where Newton's equations of motion
were integrated for $N=1000$ particles with the velocity form of the
Verlet algorithm under periodic boundary conditions and the
temperature $T$ was maintained by the Nos\'{e}-Hoover
thermostat~\cite{Frenkel1}. All the
particles have the same mass $m$. Length and time were expressed in
units of $<\sigma>$ and $\sqrt{m<\sigma>^{2}/\epsilon}$. The time
step is $\Delta t=0.002$. For convenience, the temperature $T$ in
our simulation is fixed at $k_{B}T/\epsilon=0.025$ with $k_{B}$ the
Boltzmann's constant. We use the area fraction $\phi=(1/V)\times
\sum_{j=1}^{N}\pi(\sigma_{j}/2)^{2}$ ($V=L^2$ with $L$ the box
dimension) as a control parameter. In this work, we did not make a
correction to the softness of the WCA potential and map a WCA system
to a hard-sphere system by using the effective diameter since we
only want to focus the glassy behavior as the density varies. In
athermal colloids, $1/\phi$ plays the same role as temperature in
molecular liquids. The area fraction $\phi$ in this work covers the
range from $0.62$ to $0.65$ ($L=35.01\sim35.85$). In order to check the possible finite size effects, we also performed simulations for a system with $N=4000$ in the absence of shear, and we found the qualitative agreement with the results from the smaller system. Thus, we only show the results with $N=1000$ in this paper.
At each $\phi$, the system was equilibrated for $5\times10^7$ time steps first
before collecting data and $8$ independent runs were performed.

In order to study the system under steady shear, we used the
nonequilibrium molecular dynamics (NEMD) SLLOD
algorithm~\cite{Evans, Todd}. The SLLOD equations of motion are
given by:
\begin{equation}
\begin{split}
\frac{d\textbf{r}_{j}}{dt}=\frac{\textbf{p}_{j}}{m}+\textbf{r}_{j} \cdot \nabla \textbf{v},\\
\frac{d\textbf{p}_{j}}{dt}=-\sum_{j \neq k} \frac{\partial U(\textbf{r}_{jk})}{\partial \textbf{r}_{jk}} -\textbf{p}_{j} \cdot \nabla \textbf{v},
\end{split}
\end{equation}
where $\textbf{p}_{j}$ and $\textbf{r}_{j}$ are the peculiar
momentum and laboratory position of particle $j$, and $\nabla
\textbf{v}$ is the gradient of the streaming velocity $\textbf{v}$.
Combining with the Lees-Edwards boundary conditions~\cite{Allen},
these equations of motion can generate homogeneous flow, which means
that no instability such as shear banding~\cite{Varnik1}
occurs in our simulations. In the case of planar Couette shear flow,
$\textbf{v}$ has only one non-zero component, which is given by
$\dot{\gamma}y\widehat{\textbf{x}}$, where $\dot{\gamma}$ is the
shear rate and $\widehat{\textbf{x}}$ the unit vector along the $x$
axis. Here the velocity gradient is in the $y$ direction, and the
streaming velocity in the $x$ direction. The shear rate
$\dot{\gamma}$ varies from $10^{-6}$ to $0.03$, covering the linear
and nonlinear response regimes of rheology. For each state point
under shear, $16$ independent runs were performed to improve the
statistics.

\section{Slow dynamics and growing length scales at equilibrium}

In this section, we first describe slow dynamics and growing dynamic
heterogeneity approaching the glass transition point in a quiescent
condition, and then discuss growing dynamic and static length scales
and structural features of the slow dynamics. Relation between
structure and dynamics in polydisperse colloidal liquids has also
been discussed recently in Refs.~\cite{Tanaka5} and the emergence of long-lived clusters of particles with higher order can be connected to slow dynamics and dynamic heterogeneity. Structural features of slow dynamics and dynamic heterogeneity are also confirmed by our results.

\subsection{Slow dynamics and growing dynamic heterogeneity approaching the glass transition}

We first consider the $\phi$ dependence of the potential energy $E_{p}$ and the
pressure $p$ of the system. We find that both $E_{p}$ and $p$ are
smooth functions of $\phi$ (data not shown), indicating that crystallization indeed
does not occur. Further confirmations for the avoidance of
crystallization come from the positional order and the spatial
correlation of hexatic order, which will be shown later.

\begin{figure}[!tb]
 \centering
 \includegraphics[angle=0,width=0.48\textwidth]{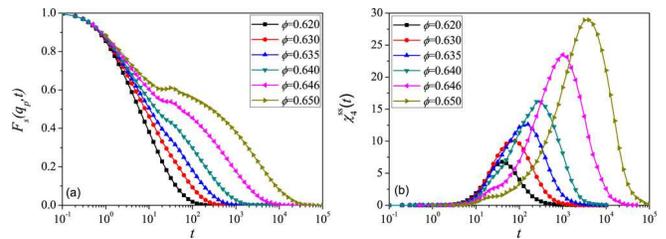}
 \caption{The self part of (a) the intermediate scattering function $F_{s}(q_{p},t)$ and (b) the four-point susceptibility $\chi_{4}^{ss}(t)$ at representative area fractions.}
\end{figure}

\begin{figure}[!tb]
 \centering
 \includegraphics[angle=0,width=0.5\textwidth]{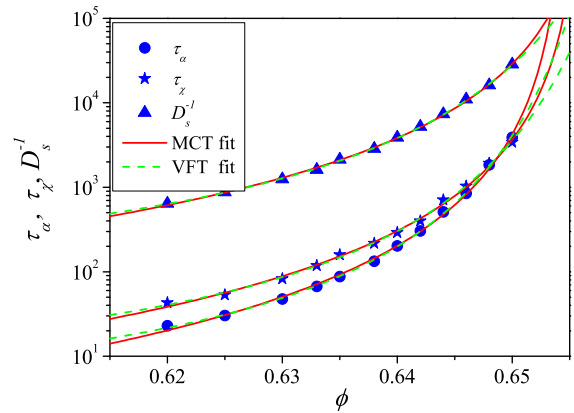}
 \caption{$\phi$-dependence of $\alpha$-relaxation time $\tau_{\alpha}$, peak time of the four-point susceptibility $\tau_{\chi}$ and self-diffusion constant $D_{s}$. The red solid lines are the results of the power-law fittings $\tau_{\alpha}\sim (\phi_{c}-\phi)^{-\gamma}$ with $\gamma=2.7$ for $\tau_{\alpha}$, $\gamma=2.6$ for $\tau_{\chi}$ and $\gamma=2.4$ for $D_{s}^{-1}$. Excellent fits are obtained to all data for a single value $\phi_{c}=0.656\pm0.001$. The green dashed lines are the results of the Vogel-Fulcher-Tamman fittings $\tau_{\alpha}\sim \exp[D\phi/(\phi_{0}-\phi)]$ with $D=0.22$ for $\tau_{\alpha}$, $D=0.23$ for $\tau_{\chi}$ and $D=0.25$ for $D_{s}^{-1}$. Excellent fits are obtained to all data for a single value $\phi_{0}=0.671\pm0.004$.}
\end{figure}

The dynamic slowing down can be characterized by calculating the
self part of the intermediate scattering function (ISF):
\begin{equation}
F_{s}(q_{p},t)=\frac{1}{N}<\sum_{j=1}^{N}\exp\{i\textbf{q}_{p}\cdot[\textbf{r}_{j}(t)-\textbf{r}_{j}(0)]\}>,
\end{equation}
where $<\cdot\cdot\cdot>$ indicates the thermal average, $i=\sqrt{-1}$ and the wave number $q_{p}$ corresponds to the
first peak of the static structure factor (which will be shown later).
$F_{s}(q_{p},t)$ can reflect structural relaxation process of a
liquid and representative results for several area fractions are
shown in Fig. 1(a). We observe that the relaxation slows down and
becomes more stretched as $\phi$ increases. In particular, a
two-step decay emerges at high enough area fractions, reflecting the
rattling motion of particles trapping within cages at short times
($\beta$-relaxation) and the motion of particles escaping from the
cages at long times ($\alpha$-relaxation). We define
$\alpha$-relaxation time as
$F_{s}(q_{p},t=\tau_{\alpha})=0.2$ (we have checked that choosing other values will not alter the qualitative results),
which is plotted as a function of $\phi$ in Fig. 2. Clearly,
$\tau_{\alpha}$ drastically increase as $\phi$ increases. For the
density range investigated, $\tau_{\alpha}$ can be well fitted by
the mode-coupling-theory (MCT) power-law~\cite{MCT}: $\tau_{\alpha}\sim
(\phi_{c}-\phi)^{-\gamma}$, where $\phi_{c}$ is the MCT glass
transition point, or the Vogel-Fulcher-Tamman (VFT) law:
$\tau_{\alpha}\sim \exp[D\phi/(\phi_{0}-\phi)]$, where $D$ is the
fragility index and $\phi_{0}$ is the ideal glass-transition point.

For the glass-forming liquids, the dynamic slowing down accompanies
growing dynamic heterogeneity. We quantify the dynamic heterogeneity
by the self part of the four-point density
correlations~\cite{Glotzer1}, which dominates results of the total four-point density
correlations~\cite{Glotzer1, Glotzer2}. A
time-dependent self-overlap order parameter $Q_{s}(t)$ is defined as
\begin{equation}
Q_{s}(t)=\frac{1}{N}<\sum_{j=1}^{N}w(|\textbf{r}_{j}(t)-\textbf{r}_{j}(0)|)>,
\end{equation}
with $w=1(0)$ for
$|\textbf{r}_{j}(t)-\textbf{r}_{j}(0)|\leq(>)0.4<\sigma>.$
The threshold $0.4$ was determined by considering the peak height of
the four-point susceptibility
$\chi_{4}^{ss}(t)=\frac{V}{N^2}[<Q_{s}(t)^{2}>-<Q_{s}(t)>^{2}]$
becoming maximal for $\phi=0.64$. $\chi_{4}^{ss}(t)$ measures the extent to which the dynamics at any two points
in space are correlated within a time interval. Typical results for $\chi_{4}^{ss}(t)$ are shown in
Fig. 1(b). As can be seen, the peak height of $\chi_{4}^{ss}(t)$
increases and the peak time $\tau_{\chi}$ shifts to larger times as
$\phi$ increases, which indicates that the dynamics not only slows
down but also becomes progressively more heterogeneous on the
approach to the glass transition. $\tau_{\chi}$ can also be well
fitted by MCT power-law or VFT law, as shown in Fig. 2. We also
present $\phi$ dependence of the self-diffusion constant $D_s$ in
Fig. 2, as evaluated by using Einstein relation from the long-time
data of the mean squared displacements. Independent fits give these
three types of data with single parameters $\phi_{c}=0.656\pm0.001$
and $\phi_{0}=0.671\pm0.004$.

\begin{figure}[!tb]
 \centering
 \includegraphics[angle=0,width=0.5\textwidth]{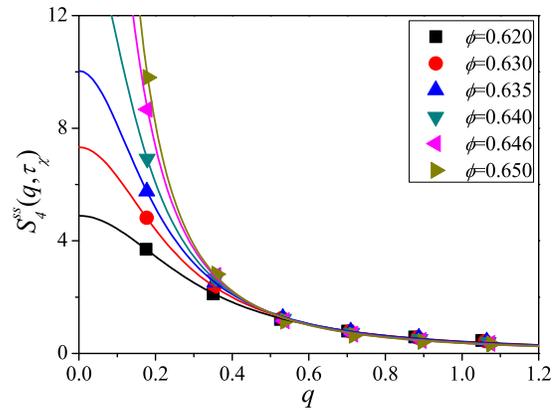}
 \caption{The structure factor of the self-overlapping particles $S_{4}^{ss}(q,t)$ at low-$q$ region for several area fractions. The solid lines are the results of the OZ fittings (see text).}
\end{figure}

As mentioned in Sec. I, the development of dynamic heterogeneity
suggests the existence of a growing dynamic length scale, which can
be investigated by the four-point, time-dependent structure factor
for the self-overlapping particles, which is defined as
\begin{equation}
S_{4}^{ss}(q,t)=\frac{L^2}{N^2}<\widetilde{\rho}(q, t)\widetilde{\rho}(-q, t)>,
\end{equation}
where
$\widetilde{\rho}(q,t)=\sum_{j=1}^{N}w(|\textbf{r}_{j}(t)-\textbf{r}_{j}(0)|)\exp[i\textbf{q}\cdot\textbf{r}_{j}(0)]$.
Here, the time $t$ is often taken as $\tau_{\chi}$, where dynamic
heterogeneity becomes most pronounced~\cite{Glotzer1}.
Representative results are given in Fig. 3.
$S_{4}^{ss}(q,\tau_{\chi})$ at low-$q$ region can be fitted by the
Ornstein-Zernike (OZ) function $S_{0}/[1+(q\xi_{4})^{2}]$ and then
we can obtain the dynamic correlation length $\xi_{4}$ in this way.
We will show the results of $\xi_{4}$ later and discuss its link to
the slow dynamics.

\subsection{Growing length scales and structural features of slow dynamics}

\begin{figure}[!tb]
 \centering
 \includegraphics[angle=0,width=0.48\textwidth]{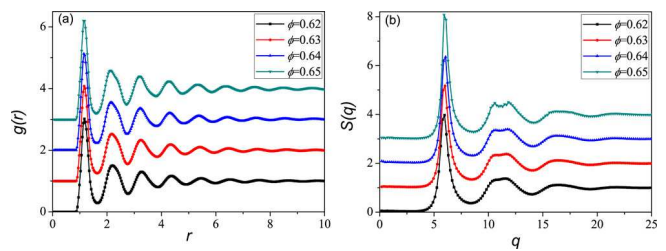}
 \caption{(a) The pair correlation function $g(r)$ and (b) the static structure factor $S(q)$ at low-$q$ region at varying $\phi$. The results have been shifted for clarity.}
\end{figure}

In Fig. 4, we present results of the pair correlation function
$g(r)$ and the static structure factor $S(q)$ at varying $\phi$, as
calculated by $g(r)=\frac{L^2}{2\pi r\Delta rN(N-1)}<\Sigma_{j \neq
k}\delta(r-|\textbf{r}_{jk}|)>$ and
$S(q)=\frac{1}{N}<\rho(q)\rho(-q)>$ respectively, where
$\rho(q)=\sum_{j=1}^{N}\exp(i\textbf{q} \cdot \textbf{r}_{j})$. On one hand, as
found in other glass-forming liquids, $g(r)$ and $S(q)$ only display
slight change on approaching the glass transition (notice that
$\tau_{\alpha}$ increases by $\sim2$ orders of magnitude in the
studied density range, as shown in Fig.2). On the other hand,
although the long-range positional order is obviously prevented from
Fig. 4, the splitting second peaks in both $g(r)$ and $S(q)$ become
more apparent as $\phi$ increases, suggesting the development of the
locally preferred order with increasing density.

\begin{figure}[!tb]
 \centering
 \includegraphics[angle=0,width=0.48\textwidth]{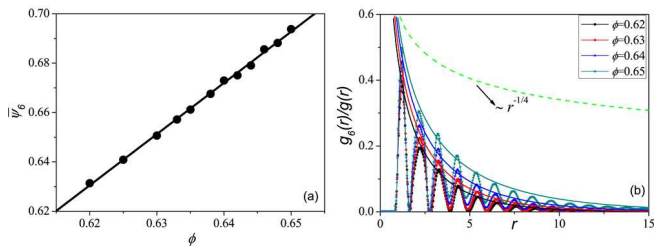}
 \caption{(a) $\phi$-dependence of $\overline{\Psi}_{6}$. The line is a linear fit to the data. (b) $\phi$-dependence of $g_{6}(r)/g(r)$. The solid lines are the results of the OZ fittings of the envelopes. The green dashed line is included to stress the fact that crystallization is avoided since $g_{6}(r)/g(r)$ decays much faster than $r^{-1/4}$. Note that $g_{6}(r)/g(r)$ decays as $r^{-1/4}$ at the boundary between the liquid phase and the hexatic phase~\cite{Nelson}.}
\end{figure}

\begin{figure}[!tb]
 \centering
 \includegraphics[angle=0,width=0.5\textwidth]{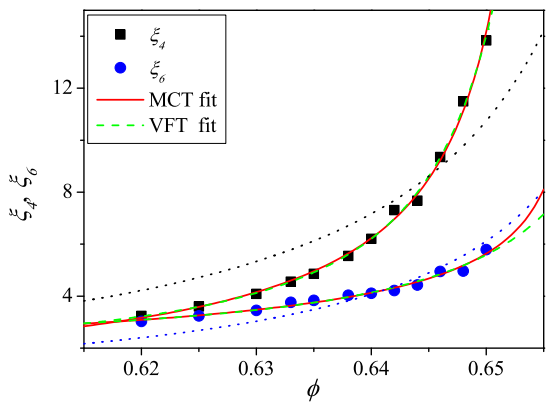}
 \caption{$\phi$-dependence of $\xi_{4}$ and $\xi_{6}$. The red solid lines are the results of the MCT fittings $\sim (\phi_{c}-\phi)^{-\gamma}$ with $\phi_{c}=0.656$ and $\gamma=0.83$ for $\xi_{4}$, $\phi_{c}=0.658$ and $\gamma=0.39$ for $\xi_{6}$. The green dashed lines are the results of the VFT fittings $\sim \exp[D\phi/(\phi_{0}-\phi)]$ with $\phi_{0}=0.670$ and $D=0.074$ for $\xi_{4}$, $\phi_{0}=0.678$ and $D=0.048$ for $\xi_{6}$. The fitting results of $\xi_{4}$ and $\xi_{6}$ for $\phi_{c}$ and $\phi_{0}$ agree well with that of $\tau_{\alpha}$, $\tau_{\chi}$ and $D_{s}$ within statistical accuracy. The dotted lines are the fittings of the functional form $\xi=\xi_{0}[\phi/(\phi_{0}-\phi)]$.}
\end{figure}

\begin{figure}[!tb]
 \centering
 \includegraphics[angle=0,width=0.5\textwidth]{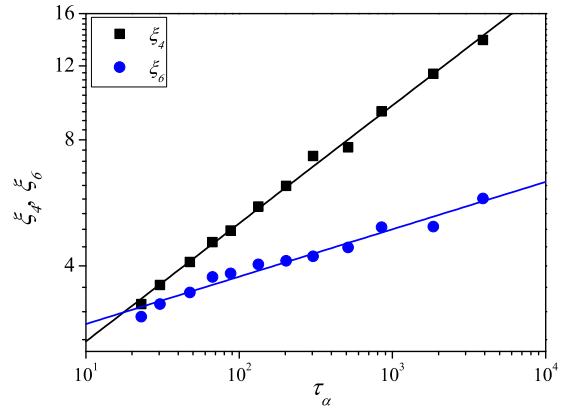}
 \caption{Log-log plot of $\xi_{4}$ and $\xi_{6}$ versus $\tau_{\alpha}$. The lines are the results of the power-law ($\xi\sim\tau_{\alpha}^{\mu}$) fittings with $\mu=0.28$ for $\xi_{4}$ and $\mu=0.11$ for $\xi_{6}$.}
\end{figure}

To characterize the local structure, we used a sixfold
bond-orientation order parameter for each particle, which is defined
as
$\psi_{6}^{j}=\frac{1}{n_{j}}\sum_{m=1}^{n_{j}}\exp(i6\theta_{m}^{j})$.
Here, $n_{j}$ is the number of the nearest neighbors for particle
$j$ which determined by the Voronoi construction, and
$\theta_{m}^{j}$ is the angle between
$(\textbf{r}_{m}-\textbf{r}_{j})$ and the $x$ axis (particle $m$ is
a neighbor of particle $j$). By construction,
$\Psi_{6}^{j}=|\psi_{6}^{j}|$ is equal to $1$ for a perfect
hexagonal arrangement of six neighbors around particle $j$ and $0$
for a random arrangement. Then, the time averaged order parameter
over $\tau_{\alpha}$ for particle $j$ is denoted by
$\overline{\Psi}_{6}^{j}=\frac{1}{\tau_{\alpha}}\int_{t_{0}}^{t_{0}+\tau_{\alpha}}\Psi_{6}^{j}dt$.
We also calculated the order parameter of the system
$\overline{\Psi}_{6}=\frac{1}{N}<\sum_{j=1}^{N}\Psi_{6}^{j}>$, which
is shown in Fig. 5(a) as a function of $\phi$. Obviously,
$\overline{\Psi}_{6}$ monotonically increases in the studied density
range as $\phi$ increases, indicating again the development of the
locally preferred order on increasing density. Here we note that
hexatic ordering in hard disc systems is a direct consequence of
dense packing and manifestation of low configurational entropy, as
emphasized in Refs.~\cite{Tanaka5}. The
spatial correlation of $\psi_{6}^{j}$ was then calculated as
$g_{6}(r)=\frac{L^2}{2\pi r\Delta rN(N-1)}<\Sigma_{j\neq
k}\delta(r-|\textbf{r}_{jk}|)\psi_{6}^{j}\psi_{6}^{k*}>$. The results
of $g_{6}(r)/g(r)$ for four different area fractions are shown in
Fig. 5(b). The envelops of $g_{6}(r)/g(r)$ can be fitted by the OZ
function $r^{-1/2}\exp(-r/\xi_{6})$, and the static correlation
length $\xi_{6}$ can be obtained in this way. We show results of
$\xi_{6}$ together with $\xi_{4}$ (as estimated from
$S_{4}^{ss}(q,t)$) in Fig. 6.  Clearly, they both increase as $\phi$
increases (or as the dynamics slows down). However, our data cannot
be well described by the functional form
$\xi=\xi_{0}[\phi/(\phi_{0}-\phi)]$ (see dotted lines in Fig. 6),
which was proposed in Refs.~\cite{Tanaka5}. Instead, they can be well fitted by MCT or VFT functions
as shown in Fig. 6, in good agreement with the results of
Ref.~\cite{Glotzer1, Glotzer2}. In fact, we find that the correlation length
scales $\xi$ have power-law dependence on $\tau_{\alpha}$ (i.e.,
$\xi\sim\tau_{\alpha}^{\mu}$), as evidenced from Fig. 7, where
log-log plot of $\xi_{4}$ and $\xi_{6}$ versus $\tau_{\alpha}$ is
presented. We will show later that this power-law relation
$\xi\sim\tau_{\alpha}^{\mu}$ holds even under shear. Thus, there are
indeed structural features of the dynamic slowing down.

\begin{figure}[!tb]
 \centering
 \includegraphics[angle=0,width=0.48\textwidth]{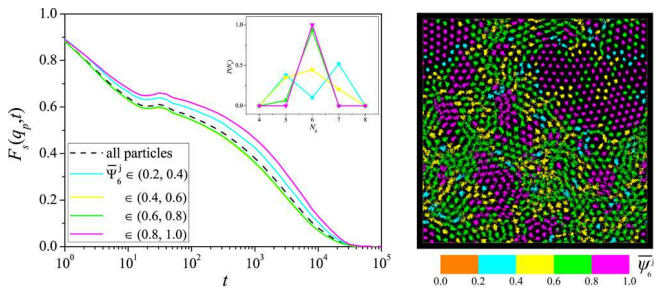}
 \caption{(Left) Main: The self part of the intermediate scattering function $F_{s}(q_{p},t)$ for particles with different structural order at $\phi=0.65$. Here, $\overline{\Psi}_{6}^{j}$ is averaged over the time window. Inset: the distribution of neighbors for particles at varying $\overline{\Psi}_{6}^{j}$. (Right) Corresponding particle trajectory during an interval of $\tau_{\alpha}$. Different colors correspond to different structural order averaged over an interval of $\tau_{\alpha}$. Clearly, particles with the highest structural order (magenta particles) are less mobile than the others and dominate the slow tail of $F_{s}(q_{p},t)$.}
\end{figure}

To further understand the relation between the slow dynamics and the
structure, we also calculated the relaxation of particles with
different structural order, as shown in Fig. 8(a). It is observed
that the particles with high structural order dominate the
relaxation process. Interestingly, the particles with
$\overline{\Psi}_{6}^{j}\in(0.2, 0.4)$ relax more slowly than those
with $\overline{\Psi}_{6}^{j}\in(0.4, 0.8)$, which is also confirmed
by the particle trajectory within a time interval of $\tau_{\alpha}$
in Fig. 8(b), i.e., the cyan particles are less mobile than the
green and yellow particles, suggesting the existence of an unexpected relationship between hexagonal structure and dynamic heterogeneity. This indicates that particles with specific structures may responsible for the slow relaxation process, which has been pointed out recently in a study of binary soft disks~\cite{Harrowell}. This feature is also evidenced by the inset of Fig. 8(a) where it reveals that the
cyan particles mainly consist of particles with five and seven neighbors.

Therefore, the structural features do exist in our model and the
slow dynamics accompanies the development of heterogeneity and
growing correlation length scales in the absence of shear. We focus on the correlation
between the slow dynamics and the structure under shear flow in the
next section.

\section{Correlation between nonequilibrium dynamics and structure under shear}

When a glass-forming liquid is sheared, marked shear-thinning behavior is expected to appear at sufficiently large shear rates. It is also found in the previous work that the shear viscosity $\eta$ and $\tau_{\alpha}$ change with shear rate $\dot{\gamma}$ with $\eta \propto\tau_{\alpha}\sim \dot{\gamma}^{-\nu}$ ($\nu>0$). However, it is not clear how correlation lengths vary with $\dot{\gamma}$ and whether there exists a connection between dynamics and structure in shear flow. In this section, we first investigate the rheological behavior and nonequilibrium dynamics of the system under steady shear and then discuss their structural features.

\begin{figure}[!tb]
 \centering
 \includegraphics[angle=0,width=0.48\textwidth]{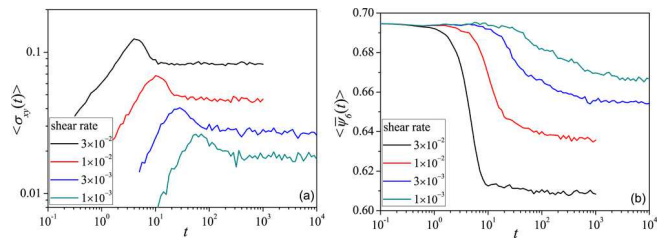}
 \caption{Time evolution of (a) the shear stress $<\sigma_{xy}>$ and (b) the bond-orientation order parameter $\overline{\Psi}_{6}$ after startup shear for several shear rates at $\phi=0.65$. The results are averaged over $200$ independent runs.}
\end{figure}

\begin{figure}[!tb]
 \centering
 \includegraphics[angle=0,width=0.48\textwidth]{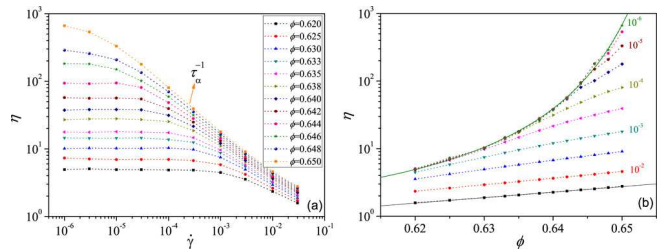}
 \caption{(a) $\dot{\gamma}$-dependence of $\eta$ at varying $\phi$. The arrow indicates that the shear-thinning behavior starts at a shear rate much slower than $1/\tau_{\alpha}$ at equilibrium ($\dot{\gamma}=0$). (b) The same data presented in the form of $\eta$ versus $\phi$ at varying $\dot{\gamma}$. The solid lines indicate that the weak $\phi$-dependence of $\eta$ at high shear rates crosses over into the strong $\phi$-dependence of $\eta$ (olive line: $\eta\sim \exp[0.2\phi/(0.667-\phi)]$) at low shear rates.}
\end{figure}

We first present the time evolution of the sample-averaged shear
stress $<\sigma_{xy}>$ after startup shear in Fig. 9(a). Here,
$\sigma_{xy}$ is defined as
\begin{equation}
\sigma_{xy}=\frac{1}{V}(\sum_{j=1}^{N}-\frac{p_{jx}p_{jy}}{m_{j}}+\sum_{j=1}^{N}\sum_{j>k}r_{jkx}\frac{\partial U(\textbf{r}_{jk})}{\partial r_{jky}}).
\end{equation}
It is seen that the system reaches the steady state after a
transient period (typically, it needs to be significantly larger
than $\dot{\gamma}^{-1}$). Before entering the steady state, a peak
emerges at some yielding shear stress. This phenomenon is usually
called overshoot behavior (or yielding behavior). Although it was
commonly found in many systems~\cite{Furukawa, Guan, Wang4}, its detailed mechanism remains elusive. Here
we show in Fig. 9(b) that overshoot behavior also accompanies
structural change in our system. The viscosity of the system was then
calculated from the steady state shear stress by using the
constitutive equation $\eta=\sigma_{xy}/\dot{\gamma}$. The shear
rate dependence of $\eta$ is given in Fig. 10(a). The same data are
also represented in the alternative form of $\eta$ versus $\phi$ in
Fig. 10(b). Clearly, the rheological behavior becomes Newtonian at
low shear rates where $\eta$ approached its equilibrium value
($\dot{\gamma}=0$), and non-Newtonian at high shear rates where
marked shear thinning ($\eta \sim \dot{\gamma}^{-\nu}$) behavior
displays. It should be noted that the shear-thinning behavior starts
at a shear rate much smaller than the inverse of $\alpha$-relaxation
time scale at equilibrium, which is consistent with previous
results~\cite{Furukawa, Yamamoto2} and indicates that there may be a
structural relaxation process much slower than $\tau_{\alpha}$
characterizing the decay of the two-body
correlation. Meanwhile, we find that
$\eta$ exhibits weak $\phi$-dependence at high shear rates and
strong $\phi$-dependence at low shear rates.

\begin{figure}[!tb]
 \centering
 \includegraphics[angle=0,width=0.48\textwidth]{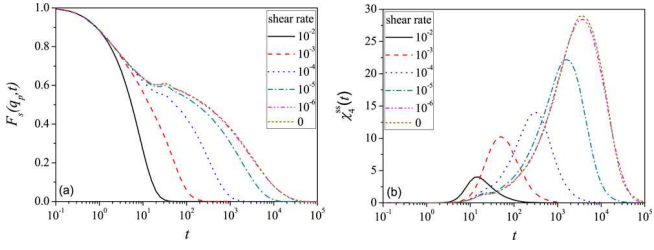}
 \caption{Shear rate dependence of (a) $F_{s}(q_{p},t)$ and (b) $\chi_{4}^{ss}(t)$ at $\phi=0.65$.}
\end{figure}

\begin{figure}[!tb]
 \centering
 \includegraphics[angle=0,width=0.48\textwidth]{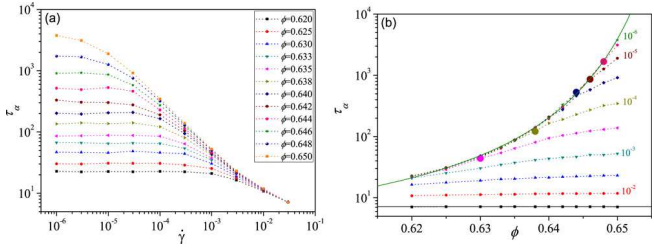}
 \caption{(a) $\dot{\gamma}$-dependence of $\tau_{\alpha}$ at varying $\phi$. (b) The same data presented in the form of $\tau_{\alpha}$ versus $\phi$ at varying $\dot{\gamma}$. The solid lines indicate that the weak $\phi$-dependence of $\tau_{\alpha}$ at high shear rates crosses over into the strong $\phi$-dependence of $\tau_{\alpha}$ (olive line: $\tau_{\alpha}\sim \exp[0.24\phi/(0.669-\phi)]$) at low shear rates. The larger symbols indicate the crossover area fractions for several shear rates which distinguish equilibrium dynamics from shear-controlled dynamics (see text).}
\end{figure}

\begin{figure}[!tb]
 \centering
 \includegraphics[angle=0,width=0.48\textwidth]{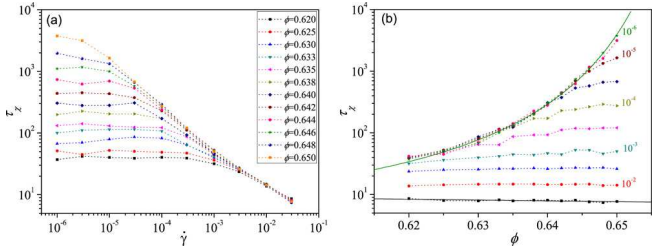}
 \caption{(a) $\dot{\gamma}$-dependence of $\tau_{\chi}$ at varying $\phi$. (b) The same data presented in the form of $\tau_{\chi}$ versus $\phi$ at varying $\dot{\gamma}$. The solid lines indicate that the weak $\phi$-dependence of $\tau_{\chi}$ at high shear rates crosses over into the strong $\phi$-dependence of $\tau_{\chi}$ (olive line: $\tau_{\chi}\sim \exp[0.25\phi/(0.671-\phi)]$) at low shear rates.}
\end{figure}

\begin{figure}[!tb]
 \centering
 \includegraphics[angle=0,width=0.48\textwidth]{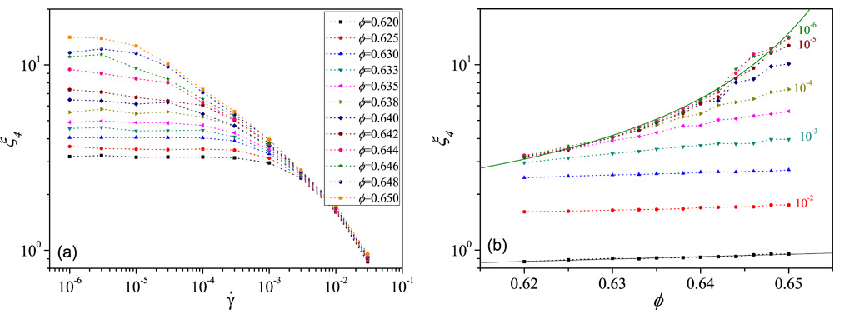}
 \caption{(a) $\dot{\gamma}$-dependence of $\xi_{4}$ at varying $\phi$. (b) The same data presented in the form of $\xi_{4}$ versus $\phi$ at varying $\dot{\gamma}$. The solid lines indicate that the weak $\phi$-dependence of $\xi_{4}$ at high shear rates crosses over into the strong $\phi$-dependence of $\xi_{4}$ (olive line: $\xi_{4}\sim \exp[0.11\phi/(0.676-\phi)]$) at low shear rates.}
\end{figure}

In order to investigate the nonequilibrium dynamics under steady
shear, Eqs. (3) and (4) are extended to the system under steady
shear, which have the following form:
\begin{equation}
F_{s}(q,t)=\frac{1}{N}<\sum_{j=1}^{N}\exp\{i[(\textbf{q}_{p}-\dot{\gamma}tq_{px}\widehat{\textbf{y}})\cdot
\textbf{r}_{j}(t)-\textbf{q}_{p} \cdot \textbf{r}_{j}(0)]\}>,
\end{equation}
and
\begin{equation}
Q_{s}(t)=\frac{1}{N}<\sum_{j=1}^{N}w(|\textbf{r}_{j}(t)-\textbf{r}_{j}(0)-\dot{\gamma}ty_{j}(0)\widehat{\textbf{x}}|)>.
\end{equation}
The results of $F_{s}(q_{p},t)$ and $\chi_{4}^{ss}(t)$ for several
shear rates at $\phi=0.65$ are given in Fig. 11. Their overall
shapes are very similar to those in quiescent conditions and the
slow decay of $F_{s}(q_{p},t)$ satisfies the time-shear
superposition property~\cite{Berthier6}. We present the results of
$\tau_{\alpha}$ and $\tau_{\chi}$ in Figs. 12 and 13. Clearly, they
share strong similarities with the viscosity: (1) $\tau_{\alpha}
\propto \tau_{\chi} \sim \dot{\gamma}^{-\nu}$ also holds in the
marked shear-thinning regime, although the exponent $\nu$ is
different for $\eta$, $\tau_{\alpha}$ and $\tau_{\chi}$; (2) $\eta$,
$\tau_{\alpha}$ and $\tau_{\chi}$ all exhibit weak $\phi$-dependence
at high $\dot{\gamma}$ and strong $\phi$ dependence at low
$\dot{\gamma}$; (3) for sufficiently low $\dot{\gamma}$, there
exists a crossover area fraction $\phi(\dot{\gamma})$ distinguishing equilibrium dynamics
from shear-controlled dynamics (see larger symbols in Fig. 12(b)). Specifically, $\eta$, $\tau_{\alpha}$ and $\tau_{\chi}$ are nearly independent on shear rate below $\phi(\dot{\gamma})$ and strongly affected by both $\phi$ and $\dot{\gamma}$ above $\phi(\dot{\gamma})$. This phenomenon is in good agreement with findings of
Ref.~\cite{Sciortino}.

We can also obtain a dynamic correlation length $\xi_{4}$ from the
four-point structure factor (Eq. (5))~\cite{comment1}, which is shown in Fig. 14.
$\xi_{4}$ also decreases in the marked shear
thinning regime as $\dot{\gamma}$ increases. However, contrary to the previous study~\cite{Heussinger1, Heussinger2, Tsamados, Nordstrom, Mizuno}, we find that power laws
fail to describe its dependence on $\dot{\gamma}$ even in the marked shear thinning regime.

\begin{figure}[!tb]
 \centering
 \includegraphics[angle=0,width=0.48\textwidth]{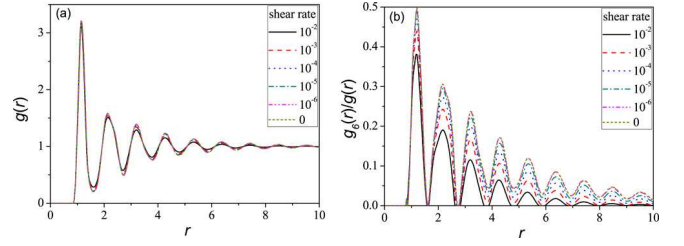}
 \caption{(a) Shear rate dependence of (a) $g(r)$ and (b) $g_{6}(r)$ at $\phi=0.65$.}
\end{figure}

\begin{figure}[!tb]
 \centering
 \includegraphics[angle=0,width=0.48\textwidth]{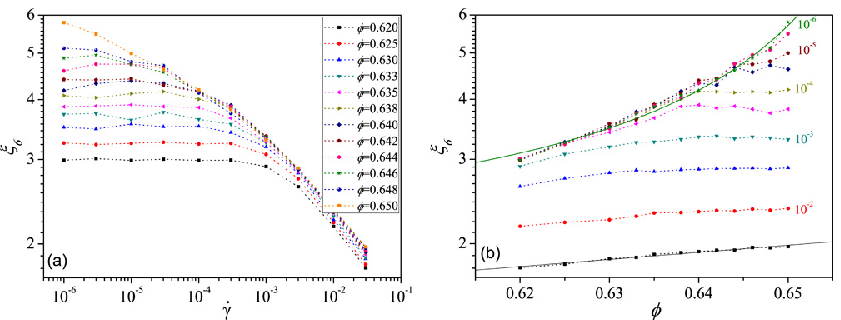}
 \caption{(a) $\dot{\gamma}$-dependence of $\xi_{6}$ at varying $\phi$. (b) The same data presented in the form of $\xi_{6}$ versus $\phi$ at varying $\dot{\gamma}$. The solid lines indicate that the weak $\phi$-dependence of $\xi_{6}$ at high shear rates crosses into the strong $\phi$-dependence of $\xi_{6}$ (olive line: $\xi_{6}\sim \exp[0.05\phi/(0.678-\phi)]$) at low shear rates.}
\end{figure}

\begin{figure}[!tb]
 \centering
 \includegraphics[angle=0,width=0.5\textwidth]{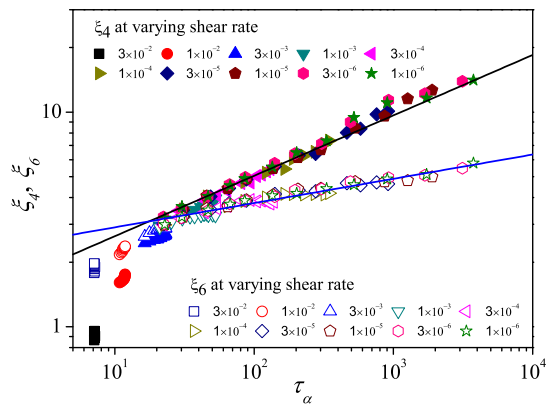}
 \caption{Log-log plot of $\xi_{4}$ and $\xi_{6}$ versus $\tau_{\alpha}$ at varying $\dot{\gamma}$. The lines are the same as those in Fig. 7.}
\end{figure}

Turing to the static properties, we present the shear rate
dependence of $g(r)$ and $g_{6}(r)$ in Fig. 15. We find that only
weak changes display in the radial distribution function (Fig.
15(a)), while pronounced changes are observed in the
bond-orientation correlation function (Fig. 15(b)). $g_{6}(r)/g(r)$
under shear can also be well fitted by the OZ function and then we
can calculate the static correlation length $\xi_{6}$ under shear,
which is shown in Fig. 16. Clearly, power laws also fail to describe
$\dot{\gamma}$ dependence of $\xi_{6}$. We also confirm that
$\xi_{4}$ and $\xi_{6}$ under shear have similar dependence on
$\dot{\gamma}$ although we didn't find an empirical or theoretical
functional form for them. However, their connection to slow dynamics
under steady shear is clear, as evidenced in Fig. 17, where we
present log-log plot of $\xi_{4}$ and $\xi_{6}$ versus
$\tau_{\alpha}$ at varying $\dot{\gamma}$. As can be seen, the
power-law relation $\xi\sim\tau_{\alpha}^{\mu}$ only breaks down for
very strong shear ($\dot{\gamma}>\sim10^{-2}$), where externally
applied forces overwhelm random thermal forces and the system
exhibit behavior like a normal liquid (e.g., the relaxation process
becomes fast and only weak dynamic heterogeneity displays under very
strong shear, see Fig. 11). However, the relation
$\xi\sim\tau_{\alpha}^{\mu}$ between $\xi_{4}$, $\xi_{6}$ and
$\tau_{\alpha}$ retains for $\dot{\gamma}<\sim10^{-3}$, where both
thermal fluctuations and applied forces are important in determining
properties of the liquids and the dynamics slows down and
accompanies strong dynamic heterogeneity. Therefore, structure
features for the slow dynamics can indeed exist even under shear
flow.

\section{Conclusions}

In conclusion, we have investigated structural features of slow
dynamics and dynamic heterogeneity in a 2D polydisperse
glass-forming liquid in both quiescent and sheared conditions. We
have confirmed the previous findings in the absence of shear that
slow dynamics accompanies the development of dynamic heterogeneity.
The correlation length scales $\xi$ grow with $\tau_{\alpha}$ as a
power-law $\xi\sim\tau_{\alpha}^{\mu}$. In the presence of shear,
both $\eta$ and $\tau_{\alpha}$ have power-law dependence on
$\dot{\gamma}$ in the marked shear thinning regime and exhibit weak
$\phi$-dependence at high shear rates and strong $\phi$-dependence
at low shear rates. The dynamic and static correlation lengths also decrease with increasing $\dot{\gamma}$, but their dependence on
$\dot{\gamma}$ cannot be described by power laws in the same $\dot{\gamma}$
regime. We further find that the relation
$\xi\sim\tau_{\alpha}^{\mu}$ between length scales and dynamics
holds for not too strong shear where thermal fluctuations and
external forces are both important in determining the properties of dense
liquids. Therefore, our results demonstrate a link between slow
dynamics and structure in glass-forming liquids even under
nonequilibrium conditions and can help to better understand the
non-Newtonian behavior of supercooled liquids and glasses.

\begin{acknowledgments}
This work is supported by the National Natural Science Foundation of
China (21074137, 50873098, 50930001, 20734003) programs and the fund
for Creative Research Groups (50921062). This work is also subsidized
by the National Basic Research Program of China (973 Program,
2012CB821500).
\end{acknowledgments}


\begin{thebibliography}{0}
\bibitem{comment1}Recent simualtions~\cite{Furukawa} indicate that the four-point structure factor $S_{4}^{ss}(q,t)$ displays strong anisotropy in the non-Newtonian regime when $t$ is significantly larger than $\tau_{\chi}$. However, isotropy of $S_{4}^{ss}(q,t)$ retains at $t\leq \tau_{\chi}$ for both Newtonian and non-Newtonian regime. Thus, we can still calculate $S_{4}^{ss}(q,\tau_{\chi})$ even in the the non-Newtonian regime and obtain the dynamic correlation length from it.
\end{thebibliography}

\end{document}